\documentclass[12pt]{amsart}
\usepackage{amssymb, amsmath}
\theoremstyle{plain}
\newtheorem{theorem}{Theorem}

\theoremstyle{definition}
\newcommand{\bi}{\bigtriangleup}

\newcommand{\ep}{\epsilon}
\newcommand{\Th}{\Theta}

\newcommand{\la}{\lambda}

\newcommand{\de}{\delta}

\newcommand{\pa}{\partial}

\newcommand{\no}{\nonumber}

\begin{document}
\title[Deformations of Dorfman's and Sokolov's Operators]
 { \sc On the Deformations of Dorfman's and Sokolov's operators\/}
\author[J.H. Chang]
{{ Jen-Hsu Chang \/}\\
  { Department of General Courses,\\
 Chung-Cheng Institute of Technology ,\\
   National Defense University,\/}\\
  { Dashi, Tau-Yuan County, Taiwan, 33509 }}
\date{\today \\
\indent e-mail: jhchang@ccit.edu.tw}
\maketitle
\begin{abstract}
 We deform the Dorfman's and Sokolov's Hamiltonian operators by the quasi-Miura transformation coming from the topological field theory and investigate the deformed operators.
\end{abstract}
\section{Introduction}
The Dorfman's and Sokolov's Hamiltonian operators are defined respectively as \cite{D1, So}($D=\partial_x$)
\begin{eqnarray}
J&=&D\frac{1}{v_x}D\frac{1}{v_x}D \label{Do} \\
S&=&v_x D^{-1} v_x \label{So},
\end{eqnarray}
which are Hamiltonian operators ( or $J^{-1}=D^{-1}v_x D^{-1}v_x D^{-1}$ and $S^{-1}= \frac{1}{v_x}D\frac{1}{v_x}$ are symplectic operators). The Dorfman's operator $J$(or $J^{-1}$) and the Sokolov's operator $S$ are related to integrable
equations as follows.
\begin{itemize}
\item The Riemann hierarchy 
\begin{eqnarray*}
v_{t_n}&=&v^n v_x=S \delta H_n=\frac{1}{(n+1)(2n+1)}K\delta H_{n+1}=\frac{1}{(n+1)(n+2)}D \delta H_{n+2}  \\
&=& \frac{1}{(n+1)(n+2)(n+3)(n+4)}J \delta H_{n+4}, 
\end{eqnarray*}
where 
\[K= Dv+vD, \quad H_n=\int v^n dx, \/ n=1,2,3 \cdots ,\]
and $\delta$ is the variational derivative. When $n=1$, it is called the Riemann equation or dispersion less KdV equation. We notice
that it seems that the Riemann hierarchy is a quater-Hamiltonian system. But one  can show that $S$ and $J$
is not compatible, i.e., $S+\lambda J$ is not Hamiltonian operator for any $\lambda \neq 0$(see below). \\
\item The Schwarzian KdV equation \cite{KN,GW}
\begin{equation}
v_t=v_{xxx}-\frac{3}{2}\frac{v_{xx}^2}{v_x}=v_x \{v,x\}=S \de H_1= J^{-1} \de H_2,
\end{equation}
where $\{v,x\}$ is the Schwartz derivative and 
\[H_1=\frac{1}{2}\int(v_x^{-2}v_{xx}^2)dx, \quad H_2=\frac{1}{2} \int (-v_x^{-2}v_{xxx}^2+\frac{3}{4}v_x^{-4}v_{xx}^4)dx.\]
\end{itemize}
{\bf Remark:} 
It is not difficult to verify that $J^{-1}$ is also a Hamiltonian operator and, then, $J$ is also a symplectic operator; however, $S^{-1}=\frac{1}{v_x}D\frac{1}{v_x}$ is \underline{not} a Hamiltonian operator and, then, $S$ is not a symplectic operator. \\

\indent Next, to deform the operators $J$  and $S$ , we use the free energy in topological field theory
of the famous KdV equation
\begin{equation}
u_t=uu_x+ \frac{\ep^2}{12}u_{xxx} \label{KdV}
\end{equation}
to construct the quasi-Miura transformation as follows. The free energy $F$ of  KdV equation \eqref{KdV} in TFT 
has the form($F_0= \frac{1}{6}v^3$)
\[F=\frac{1}{6}v^3+\sum_{g=1}^{\infty} \ep^{2g-2} F_g(v;v_x, v_{xx}, v_{xxx}, \cdots, v^{(3g-2)}).\]
Let 
\begin{eqnarray*}
\bi F &=& \sum_{g=1}^{\infty} \epsilon^{2g-2} F_g(v;v_x, v_{xx}, v_{xxx}, \cdots, v^{(3g-2)}) \\
                 &=& F_1(v;v_x)+ \ep^2 F_2(v;v_x, v_{xx},v_{xxx},v_{xxxx}) \\
                  &+&\ep^4 F_3(v;v_x, v_{xx},v_{xxx},v_{xxxx},\cdots,v^{(7)})+\cdots .
\end{eqnarray*}
The $\bi F$ will satisfy the loop equation(p.151 in \cite{DZ})
\begin{eqnarray}
&&\sum_{r \geq 0}\frac{\pa \bi F}{\pa v^{(r)}} \pa_x^r \frac{1}{v-\lambda}+\sum_{r \geq 1} \frac{\pa \bi F}{\pa v^{(r)}} \sum_{k=1}^r \left(\begin{array}{c} r \\k \end{array} \right) \label{loop} \\
 &&\pa_x^{k-1} \frac{1}{\sqrt{v-\lambda}} \pa_x^{r-k+1} \frac{1}{\sqrt{v-\lambda}} \no \\
&=&\frac{1}{16 \lambda^2}-\frac{1}{16(v-\lambda)^2}-\frac{\kappa_0}{\lambda^2} \no \\
&+&\frac{\ep^2}{2}\sum_{k,l \geq 0}\left[\frac{\pa^2 \bi F}{\pa v^{(k)}\pa v^{(l)}}+\frac{\pa \bi F}{\pa v^{(k)}}\frac{\pa \bi F}{\pa v^{(l)}}\right] \pa_x^{k+1} \frac{1}{\sqrt{v-\lambda}}\pa_x^{l+1} \frac{1}{\sqrt{v-\lambda}} \no \\
&-&\frac{\ep^2}{16}\sum_{k \geq 0}\frac{\pa \bi F}{\pa v^{(k)}}\pa_x^{k+2} \frac{1}{(v-\lambda)^2}. \no
\end{eqnarray}  
Then we can determine $F_1, F_2, F_3, \cdots$ recursively by substituting $\bi F$ into equation \eqref{loop}. For $F_1$, one obtains 
\[\frac{1}{v-\lambda}\frac{\pa F_1}{\pa v}-\frac{3}{2}\frac{v_x}{(v-\lambda)^2}\frac{\pa F_1}{\pa v_x}=\frac{1}{16 \lambda^2}-\frac{1}{16 (v-\lambda)^2}-\frac{\kappa_0}{\lambda^2}.\]
From this, we have 
\[\kappa_0=\frac{1}{16}, \quad F_1=\frac{1}{24} \log v_x.\]
For the next terms $F_2(v;v_x, v_{xx},v_{xxx},v_{xxxx})$, it can be similarly computed and the result is 
\[F_2=\frac{v_{xxxx}}{1152v_x^2}-\frac{7v_{xx}v_{xxx}}{1920v_x^3}+\frac{v_{xx}^3}{360v_x^4}.\]
Now, one can define the quasi-Miura transformation as
\begin{eqnarray}
\label{qua}\\
 u&=&v+\ep^2 (\bi F)_{xx}=v+ \ep^2(F_1)_{xx}+ \ep^4(F_2)_{xx}+\cdots \no \\
 &=&v+\frac{\ep^2}{24}(\log v_x)_{xx}+ \ep^4(\frac{v_{xxxx}}{1152v_x^2}-\frac{7v_{xx}v_{xxx}}{1920v_x^3}+\frac{v_{xx}^3}{360v_x^4})_{xx}+ \cdots. \no
\end{eqnarray}
One remarks that Miura-type transformation means the coefficients of $\ep$ are homogeneous polynomials in the derivatives $v_x, v_{xx}, \cdots , v^{(m)}$(p.37 in \cite{DZ}, \cite{LO}) and "quasi" means the ones of $\ep$ are quasi-homogeneous rational functions in the derivatives, too (p.109 in \cite{DZ} and see also \cite{Sr} ). \\
\indent The truncated quasi-Miura transformation 
\begin{equation}
u=v+\sum_{n=1}^g \ep^{2n}\left[F_n(v;v_x, v_{xx}, \cdots, v^{(3g-2)}) \right]_{xx} \label{tqua}
\end{equation}
has the basic property (p.117 in \cite{DZ}) that it reduces the Magri Poisson pencil \cite{MA} of KdV equation \eqref{KdV} 
\begin{equation}
\{u(x), u(y)\}_{\la}=[u(x)-\la]D\de(x-y)+\frac{1}{2}u_x(x)\de (x-y)+\frac{\ep^2}{8}D^3 \de(x-y) \label{Mag}
\end{equation}
to the Poisson pencil of  the Riemann hierarchy \eqref{Rie}:
\begin{equation}
\{v(x), v(y)\}_{\la}=[v(x)-\la]D\de(x-y)+\frac{1}{2}v_x(x)\de (x-y)+ O(\ep^{2g+2}).\label{Mag1}
\end{equation}
One can also say that the truncated quasi-Miura transformation  \eqref{tqua} deforms the KdV equation \eqref{KdV} to the Riemann equation $v_t=vv_x$ up to $O(\ep^{2g+2})$. \\
{\bf Remark:}
 A simple calculation shows that, under the transformation $u=\frac{\ep^2}{4} \{m,x\}$, the KdV equation \eqref{KdV} is transformed into the Schwarzian KdV equation
\[m_t=\frac{\ep^2}{12}m_x \{m,x\}=\frac{\ep^2}{12}(m_{xxx}-\frac{3}{2}\frac{m_{xx}^2}{m_x}). \]
Furthermore, after a direct calculation, one can see that the Magri Poisson bracket
\begin{equation}
K(\ep)=\{u(x), u(y)\}=u(x)D\de(x-y)+\frac{1}{2}u_x(x)\de (x-y)+\frac{\ep^2}{8}D^3 \de(x-y) \label{MP}
\end{equation}
is transformed into the Dorfman's symplectic  operator $J^{-1}$  ($m=v$)
\[\{m(x),m(y)\}=-\frac{\ep^2}{8}D^{-1}m_xD^{-1}m_xD^{-1} \de (x-y).\]
\indent Now, a natural question arises: under the truncated quasi-Miura transformation \eqref{tqua}, are the deformed
Dorfman's operator $J(\ep)$ and Sokolov's operator $S(\ep)$ still Hamiltonian operators up to $O(\ep^{2g+2})$? For simplicity, we consider 
only the case $g=1$, i.e., 
\begin{equation}
u=v+ \frac{\ep^2}{24}(\log v_x)_{xx}+O(\ep^4) \label{qua1}
\end{equation}
or
\begin{equation}
v=u - \frac{\ep^2}{24}(\log u_x)_{xx}+O(\ep^4). \label{qua2}
\end{equation}
The answer is true for the Dorfman's operator $J(\ep)$  but it's \underline{false} for the Sokolov's  operator $S(\ep)$. It's the purpose of this article.
\section{Deformations under quasi-Miura Transformation}
In the new "u-coordinate", $J$ and $S$ will be given by the operators
\begin{eqnarray}
J(\ep)&=&M^{*}D\frac{1}{u_x - \frac{\ep^2}{24}(\log u_x)_{xxx}}D\frac{1}{u_x - \frac{\ep^2}{24}(\log u_x)_{xxx}}DM \label{dej} \\
&+&O(\ep^4); \no  \\
S(\ep)&=&M^{*}(u_x - \frac{\ep^2}{24}(\log u_x)_{xxx})D^{-1}(u_x - \frac{\ep^2}{24}(\log u_x)_{xxx})M  \label{des} \\
&+&O(\ep^4), \no 
\end{eqnarray}
where
\begin{eqnarray*}
M &=& 1-\frac{\ep^2}{24}D \frac{1}{u_x}D^2 \\
M^* &=& 1+\frac{\ep^2}{24}D^2 \frac{1}{u_x}D,
\end{eqnarray*}
$M^*$ being the adjoint operator of $M$. Then we have the following 
\begin{theorem}
(1)$J(\ep)$ is a Hamiltonian operator up to $O(\ep^4).$ (2)$S(\ep)$ is \underline{not} a Hamiltonian operator up to $O(\ep^4)$.
\end {theorem}
\begin{proof}(1)The fact that $J(\ep)$ is a skew-adjoint (or $J^*(\ep)=-J(\ep)$) differential operator (up to $O(\ep^4)$)
follows immediately from \eqref{dej}. Rather than prove the Poisson form \cite{Pe} of the Jacobi identity for $J(\ep)$, it is simpler to prove that the symplectic two form 
\[ \Omega_J(\ep)=\int \{du \wedge J(\ep)^{-1} du \}dx +O(\ep^4) \]
is closed \cite{ON,O3}: $d\Omega_J(\ep)=O(\ep^4)$. \\
\indent A simple calculation can yield 
\begin{eqnarray*}
J(\ep)^{-1} &=&(1+\frac{\ep^2}{24}D \frac{1}{u_x}D^2)D^{-1}(u_x - \frac{\ep^2}{24}(\log u_x)_{xxx})D^{-1}(u_x - \frac{\ep^2}{24}(\log u_x)_{xxx})D^{-1} \\
&&(1-\frac{\ep^2}{24}D^2 \frac{1}{u_x}D) \\
&=&(D^{-1}u_x-\frac{\ep^2}{24}D^{-1}(\log u_x)_{xxx}+\frac{\ep^2}{24}D\frac{1}{u_x}Du_x)D^{-1}(u_x D^{-1}-\frac{\ep^2}{24}(\log u_x)_{xxx}D^{-1}\\
&-&\frac{\ep^2}{24}u_xD\frac{1}{u_x}D)+O(\ep^4) \\
&=&D^{-1}u_xD^{-1}u_xD^{-1}+\frac{\ep^2}{24} [D\frac{1}{u_x}Du_xD^{-1}u_xD^{-1}-D^{-1}(\log u_x)_{xxx}D^{-1}u_xD^{-1}\\
&-&D^{-1}u_xD^{-1}u_xD\frac{1}{u_x}D-D^{-1}u_xD^{-1}(\log u_x)_{xxx}D^{-1}]+O(\ep^4)\\
&=&D^{-1}u_xD^{-1}u_xD^{-1}+\frac{\ep^2}{24}[Du_xD^{-1}-D^{-1}u_xD+(\log u_x)_xu_xD^{-1}\\
&+&D^{-1}(\log u_x)_xu_x]+O(\ep^4).
\end{eqnarray*}
Let $\psi$ denote the potential function  for $u$, i.e., $u=\psi_x$. Thus, formally, 
\[D_x^{-1}(du)=d \psi \]
and hence, after a series of integration by parts, one has 
\begin{eqnarray*}
\Omega_J(\ep) &=& \int \{[(D^{-1}d(\frac{\psi_x^2}{2})) \wedge d(\frac{\psi_x^2}{2})-\psi_x d \psi \wedge d(\frac{\psi_x^2}{2})] \\
&+&\frac{\ep^2}{24}[2\psi_{xx} d \psi \wedge d \psi_{xx} +2\psi_{xxx} d \psi_x \wedge d \psi]\}dx+O(\ep^4).
\end{eqnarray*}
So 
\begin{eqnarray*}
d\Omega_J(\ep) &=& \int \{0+\frac{\ep^2}{12}[d \psi_{xxx} \wedge d\psi_x \wedge d \psi]\} dx+O(\ep^4)\\
&=&\frac{\ep^2}{12}\int \{(d \psi_{xx} \wedge d\psi_x \wedge d \psi)_x\} dx+O(\ep^4)=O(\ep^4).
\end{eqnarray*}
This completes the proof of (1). \\
(2)The skew-adjoint property of the deformed Sokolov's operator $S(\ep)$ \eqref{des} is obvious. To see whether $S(\ep)$
is Hamiltonian operator or not, we must check $S(\ep)$ whether satisfy the Jacobi identity up to $O(\ep^4)$. Following \cite{Pe,ON}, we introduce the arbitrary basis of tangent vector $\Th$,  which is then conveniently manipulated 
according to the rules of exterior calculus. The Jacobi identity is given by the compact expression 
\begin{equation}
P(\ep) \wedge \de I= O(\ep^4) \,(mod. \, div.), \label{Van} 
\end{equation}
where $P(\ep)=S(\ep)\Th$, \, $I=\frac{1}{2} \Th \wedge P(\ep)$ and $\de$ denotes
the variational derivative. The vanishing of the tri-vector \eqref{Van} modulo
a divergence is equivalent to the satisfication of the Jacobi identity.\\
After a tedious calculation, one can obtain
\begin{eqnarray*}
S(\ep)&=& M^* (u_x - \frac{\ep^2}{24}(\log u_x)_{xxx})D^{-1}(u_x - \frac{\ep^2}{24}(\log u_x)_{xxx})M+O(\ep^4) \\
&=& [u_x+\frac{\ep^2}{24}(D^3+D^2(\log u_x)_x-(\log u_x)_{xxx})]D^{-1}[u_x-\frac{\ep^2}{24}(D^3-(\log u_x)_xD^2 \\
&+&(\log u_x)_{xxx})]+O(\ep^4) \\
&=&u_x D^{-1}u_x+\frac{\ep^2}{24}[D^2u_x+D^2(\log u_x)_x D^{-1}u_x-(\log u_x)_{xxx}D^{-1}u_x-u_xD^2\\
&+&u_xD^{-1}(\log u_x)_x D^2-u_xD^{-1}(\log u_x)_{xxx}]+O(\ep^4) \\
&=&u_x D^{-1}u_x+\frac{\ep^2}{24}[D^2u_x-u_xD^2+(\log u_x)_xDu_x+u_xD(\log u_x)_x]+O(\ep^4)\\
&=&u_x D^{-1}u_x+\frac{\ep^2}{12}[Du_{xx}+u_{xx}D]+O(\ep^4).
\end{eqnarray*}
So
\[P(\ep)=S(\ep)\Th=u_xD^{-1}(u_x\Th)+\frac{\ep^2}{12}[2u_{xx}\Th_x+u_{xxx}\Th]+O(\ep^4).\]
Hence 
\[I=\frac{1}{2}\Th \wedge P(\ep)=\frac{1}{2}u_x \Th \wedge D^{-1}(u_x \Th)
+\frac{\ep^2}{12}u_{xx}\Th \wedge \Th_x+O(\ep^4)\]
and then 
\begin{eqnarray*}
\de I &=&-\frac{1}{2}[\Th \wedge D^{-1}(u_x \Th)]_x-\frac{1}{2}u_x \Th \wedge D^{-1}(\Th_x)+\frac{\ep^2}{12}[\Th \wedge \Th_x]_{xx}+O(\ep^4) \\
&=&-\frac{1}{2}\Th_x \wedge D^{-1}(u_x \Th)+\frac{\ep^2}{12}[\Th \wedge \Th_x]_{xx}+O(\ep^4). 
\end{eqnarray*}
Finally, 
\begin{eqnarray*}
P(\ep) \wedge \de I &=&\{u_xD^{-1}(u_x\Th)+\frac{\ep^2}{12}[2u_{xx}\Th_x+u_{xxx}\Th] \} \wedge \{-\frac{1}{2}\Th_x \wedge D^{-1}(u_x \Th)\\
&+&\frac{\ep^2}{12}[\Th \wedge \Th_x]_{xx} \}+O(\ep^4) \\
&=&0+\frac{\ep^2}{12}\{-\frac{1}{2}u_{xxx}\Th \wedge \Th_x \wedge D^{-1}(u_x \Th)+u_{xxx}D^{-1}(u_x \Th) \wedge
\Th \wedge \Th_x \\
&+&3u_{xx}u_x \Th \wedge \Th \wedge \Th_x+u_x^2 \Th_x \wedge \Th \wedge \Th_x \}+O(\ep^4) \\
&=&0+\frac{\ep^2}{24}u_{xxx}\Th \wedge \Th_x \wedge D^{-1}(u_x \Th),
\end{eqnarray*}
which can be easily checked that it can't be expressed as a total divergence. So $S(\ep)$ can't satisfy the Jacobi identity
and therefore $S(\ep)$ is not a Hamiltonian operator. This completes the proof of (2).
\end{proof}
{\bf Remark:}
Using the technics of the last proof, one can show that $J$ and $S$ is \underline{not} compatible.  Since $J$ and $S$ are Hamiltonian operators, what we are going to do is show that \cite{Pe, ON}
\[\tilde Q (\Th)\wedge \de R +Q(\Th) \wedge \de \tilde R \neq 0,  \quad (mod.\quad div.) \]
where 
\begin{eqnarray*}
Q(\Th)&=& v_x D^{-1}(v_x \Th), \quad R=\frac{1}{2} \Th \wedge Q(\Th) \\
\tilde Q(\Th) &=&(\frac{1}{v_x}(\frac{\Th_x}{v_x})_x)_x, \quad \tilde R=\frac{1}{2} \Th \wedge \tilde Q(\Th)=-\frac{1}{2v_x^2} \Th_x \wedge \Th_{xx}.
\end{eqnarray*}
Then 
\[\de R=\frac{-1}{2}[\Th \wedge D^{-1}(v_x \Th)]_x-\frac{1}{2}v_x\Th \wedge D^{-1}(\Th_x)=\frac{-1}{2}\Th_x \wedge D^{-1}(v_x \Th)\]
and
\[\de \tilde R=-(\frac{1}{v_x^3} \Th_x \wedge \Th_{xx})_x. \]
Hence
\begin{eqnarray*}
&&\tilde Q (\Th)\wedge \de R +Q(\Th) \wedge \de \tilde R \\
&=&(\frac{1}{v_x}(\frac{\Th_x}{v_x})_x)_x \wedge (\frac{-1}{2}\Th_x \wedge D^{-1}(v_x \Th))-v_x D^{-1}(v_x \Th) \wedge
(\frac{1}{v_x^3} \Th_x \wedge \Th_{xx})_x \\
&=&\frac{1}{2}\frac{1}{v_x}(\frac{\Th_x}{v_x})_x \wedge [\Th_{xx}\wedge D^{-1}(v_x \Th)+v_x \Th_x \wedge \Th]+[v_{xx}
D^{-1}(v_x \Th)+v_x^2 \Th] \\
&\wedge&(\frac{1}{v_x^3} \Th_x \wedge \Th_{xx}) \\
&=&\frac{1}{2v_x} \Th_{xx}\wedge \Th_x \wedge \Th-\frac{v_{xx}}{2v_x^3} \Th_{x}\wedge \Th_{xx} \wedge D^{-1}(v_x \Th)
+  \frac{v_{xx}}{v_x^3} D^{-1}(v_x \Th) \wedge \Th_x \wedge \Th_{xx}  \\
&+&\frac{1}{v_x} \Th \wedge \Th_x \wedge \Th_{xx} \\
&=&\frac{1}{2v_x} \Th \wedge \Th_x \wedge \Th_{xx}+\frac{v_{xx}}{2v_x^3} \Th_{x}\wedge \Th_{xx} \wedge D^{-1}(v_x \Th)\\
&\neq 0& \quad (mod.\quad div.),
\end{eqnarray*}
as required.
\section{Concluding Remarks}
\begin{itemize}
\item That $J(\ep)$ is a Hamiltonian operator (up to $O(\ep^4)$) is proved in \cite{CH}. One gives another proof here, which 
remarkably simplifies the proof given in \cite{CH}.
\item We notice that all the deformed operators $J(\ep)$\eqref{dej}, $D(\ep)(=D+O(\ep^4))$, $K(\ep)$\eqref{MP} under 
the quasi-Miura transformation \eqref{qua} are Hamiltonian operators(up to $O(\ep^4)$). That the deformed Sokolov's 
operator $S(\ep)$ is not Hamiltonian is a little surprised, which means that the Poisson bracket of the Hamiltonians
$H_m(u;\ep)$, $H_n(u;\ep)$ for $S(\ep)$
\[\{H_m(u;\ep), H_n(u;\ep) \}_{S(\ep)}\]
won't be $O(\ep^4)$ but $O(\ep^2)$, i.e., it can't be a conserved quantity of the Riemann hierarchy .
\end{itemize}

\subsection*{Acknowledgments}
The author thanks for the support of National Science Council under grant no. NSC 91-2115-M-014-001 .

\end{document}